\documentclass{article}

\usepackage{microtype}
\usepackage{graphicx}

\usepackage{subfig}
\usepackage{booktabs} % for professional tables

\usepackage{amsfonts}
\usepackage{textcomp}
\usepackage{xcolor}
\usepackage{multirow}
\usepackage{url}
\usepackage{array}

\usepackage{hyperref}

\usepackage[accepted]{icml2025}

\usepackage{amsmath}
\usepackage{amssymb}
\usepackage{mathtools}
\usepackage{amsthm}

\usepackage[capitalize,noabbrev]{cleveref}

\theoremstyle{plain}

\theoremstyle{definition}

\theoremstyle{remark}

\usepackage[textsize=tiny]{todonotes}

\icmltitlerunning{Efficient and Robust Semantic Image Communication via Stable Cascade}

\begin{document}

\twocolumn[
\icmltitle{Efficient and Robust Semantic Image Communication via Stable Cascade}



\icmlsetsymbol{equal}{*}

\begin{icmlauthorlist}
\icmlauthor{Bilal Khalid}{yyy}
\icmlauthor{Pedro Freire}{yyy}
\icmlauthor{Sergei K. Turitsyn}{yyy}
\icmlauthor{Jaroslaw E. Prilepsky}{yyy}

%\icmlauthor{Bilal Khalid}{equal,yyy}
%\icmlauthor{Pedro Freire}{equal,yyy,comp}
%\icmlauthor{Firstname3 Lastname3}{comp}
%\icmlauthor{Firstname4 Lastname4}{sch}
%\icmlauthor{Firstname5 Lastname5}{yyy}
%\icmlauthor{Firstname6 Lastname6}{sch,yyy,comp}
%\icmlauthor{Firstname7 Lastname7}{comp}

%\icmlauthor{Firstname8 Lastname8}{sch}
%\icmlauthor{Firstname8 Lastname8}{yyy,comp}
\end{icmlauthorlist}

\icmlaffiliation{yyy}{Aston Institute of Photonic Technologies, Aston University, Birmingham, UK}

\icmlcorrespondingauthor{Bilal Khalid}{r.khalid4@aston.ac.uk}

\icmlkeywords{Generative AI, Semantic Communication, Latent Diffusion Model, Image Transmission}

\vskip 0.3in
]



\printAffiliationsAndNotice{} % otherwise use the standard text.

\begin{abstract}
Diffusion Model (DM) based Semantic Image Communication (SIC) systems face significant challenges, such as slow inference speed and generation randomness, that limit their reliability and practicality. To overcome these issues, we propose a novel SIC framework inspired by Stable Cascade, where extremely compact latent image embeddings are used as conditioning to the diffusion process. Our approach drastically reduces the data transmission overhead, compressing the transmitted embedding to just $0.29\%$ of the original image size. It outperforms three benchmark approaches --- the diffusion SIC model conditioned on segmentation maps (GESCO), the recent Stable Diffusion (SD)-based SIC framework (Img2Img-SC), and the conventional JPEG2000 $+$ LDPC coding --- by achieving superior reconstruction quality under noisy channel conditions, as validated across multiple metrics. Notably, it also delivers significant computational efficiency, enabling over $3\times$ faster reconstruction for $512\times512$ images and more than $16\times$ faster for $1024\times1024$ images as compared to the approach adopted in Img2Img-SC. 
\end{abstract}

\begin{figure}[t]
    \begin{flushright}
    \includegraphics[width=\columnwidth]{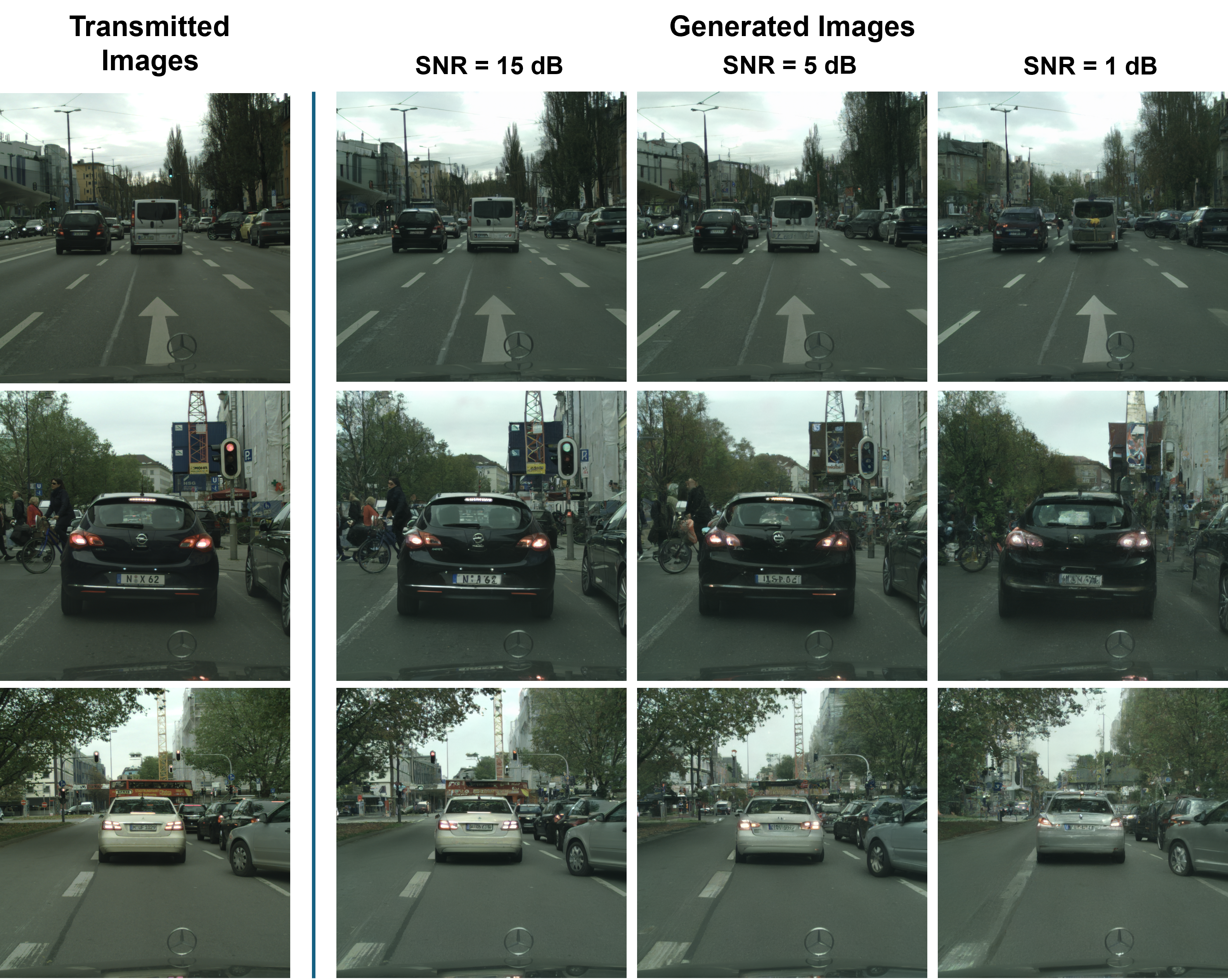}
    \vspace{-20pt}
    \end{flushright}
    \caption{$1024\times1024$ Image reconstructions using our model under different channel SNR conditions. Even at an SNR of 1 dB, images are faithfully reconstructed and perceptually very similar to the transmitted images.}
    \label{fig:Reconstructions}
    %\vspace{-15pt}
    \vskip -0.15in
\end{figure}

\section{Introduction}
\label{intro}

Semantic communication (SemCom) is a transformative approach that focuses on effectively conveying the meaning of information rather than transmitting raw bit data~\cite{strinati20216g}. The goal is to communicate the essential information the receiver needs to complete its task successfully. This also makes it bandwidth efficient as significantly less data has to be transmitted across the communication channel \cite{luo2022semantic, qin2021semantic}.

\indent The advancement of Deep Learning (DL) and generative AI has enabled the emergence of SemCom as a viable alternative to traditional communication. DL and generative AI models are used for extracting the relevant semantic information at the transmitter end as well as for deciphering the meaning behind this information at the receiver end. Deep learning-based Joint Source-Channel Coding (DeepJSCC) \cite{bourtsoulatze2019deep} was one of the first approaches to incorporate DL in wireless system design. Variational Autoencoders (VAEs), Generative Adversarial Networks (GANs), Diffusion Models (DMs) and Flow-based Generative Models (FGMs) are the major generative AI techniques now commonly used in SemCom systems \cite{xia2023generative}. Out of these, DMs have shown great potential at Semantic Image Communication (SIC) tasks because of their exceptional ability to synthesize high-quality images~\cite{dhariwal2021diffusion}. However, one drawback of DMs is that they are inherently slower at inference because of their iterative nature. The introduction of Latent Diffusion Models (LDMs)~\cite{Rombach_2022_CVPR} has alleviated this problem by performing the diffusion process in a compressed latent space instead of the original pixel space, enabling fast and high-resolution image generation via diffusion.

\begin{figure*}[t]
    \centering
    \includegraphics[width=\textwidth]{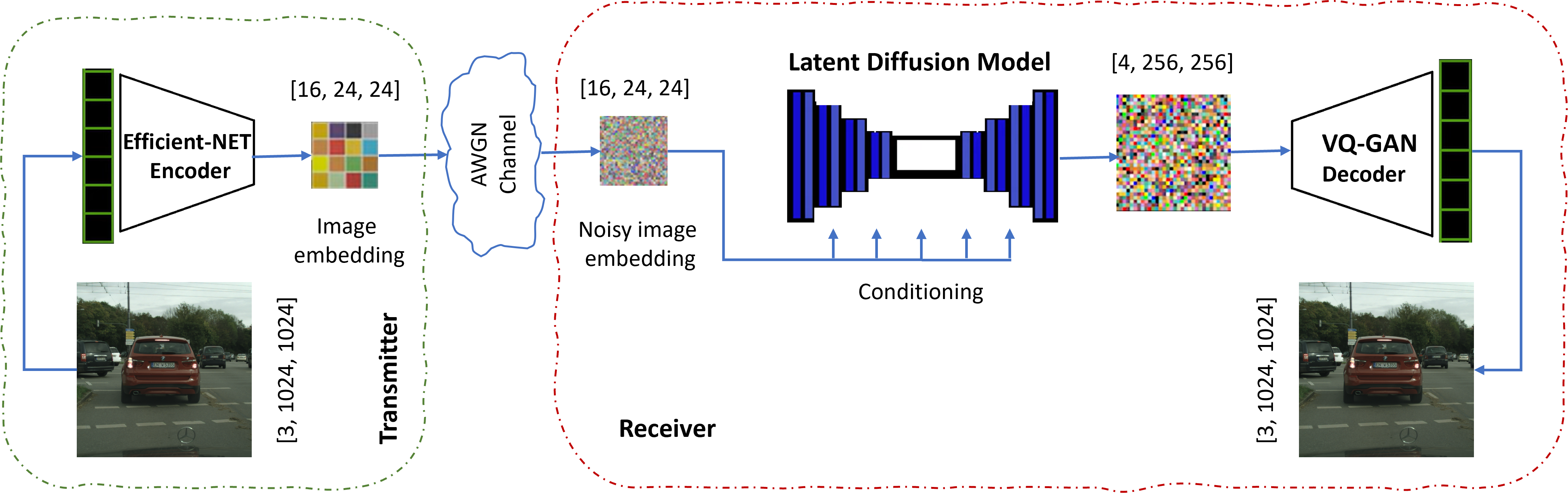}
    \caption{Our system model. At the transmitter side, a compact image embedding $Z$ of size [16, 24, 24] is extracted from an image $X$ of size [3, 1024, 1024]. $Z$ is transmitted across the physical channel. The receiver uses the noisy embedding $\hat{Z}$ as conditioning for the LDM. Finally, the VQGAN decoder is used to project the image back into pixel space.}
    \label{fig:System_Model}
    %\vspace{-10pt}
    \vskip -0.1in
\end{figure*}

\indent Several DM-based SIC systems have been implemented in recent years. In \cite{grassucci2023generative}, segmentation maps are used to guide the diffusion process. In \cite{yilmaz2024high}, the primary image structure is transmitted using the DeepJSCC technique, whereas fine details are generated using the diffusion model. \cite{Jiang2024} also use a diffusion model to refine the reconstruction obtained after image decoding. However, inference using these approaches is time-consuming. Recently, LDMs have been used for SIC to speed up the inference process. In \cite{nam2024language, cicchetti2024language}, text conditioning is used to guide the generative process of Stable Diffusion's text-to-image model \cite{Rombach_2022_CVPR}. In \cite{cicchetti2024language}, the generation process starts from a noisy version of image embedding instead of pure noise. Although efficient in terms of bandwidth, these models struggle to faithfully reconstruct the intended image and suffer from generation randomness. \cite{chen2024casc} denoise a noisy image embedding using an LDM, and the clean embedding is then used to reconstruct the image using a semantic decoder. Instead of predicting the noise in the image, \cite{yang2024rate} use a diffusion model to predict the source image in a few denoising steps directly. Both of these models reduce inference time but operate at a lower compression factor as compared to our proposed method. 

\indent In this paper, we propose a novel SIC model inspired by Stable Cascade (SC) \cite{pernias2023wurstchen}, a multistage text-to-image LDM that operates in a much smaller latent space than Stable Diffusion (SD). Our approach achieves the trifecta of high compression efficiency, fast inference, and perceptually aligned image reconstruction, which is missing in existing DM-based SIC systems. In our method, a highly compressed image embedding is extracted using a semantic encoder and transmitted across the physical channel. The noisy embedding is then given as a conditioning signal to the LDM of SC that projects it into a higher dimensional latent space where the semantic decoder operates. Results indicate that we outperform benchmark models and as shown in \cref{fig:Reconstructions}, generate consistent reconstructions even under extremely poor channel Signal-to-Noise Ratio (SNR) conditions.

\section{Proposed Framework}

In this section, the proposed system model is explained. The model is built upon the architecture of SC that has three stages, i.e., stages A, B and C. As discussed below, our model is based on stage A and a finetuned stage B that is trained to work with noisy conditioning.

Stage A is a Vector Quantized Generative Adversarial Network (VQGAN) \cite{esser2021taming} with parameters $\Theta$ that compresses the image space by a factor of $4$. The relationship between an input image $X \in \mathbb{R}^{3 \times 1024 \times 1024}$ and the output of VQGAN encoder $X_{\text{VG}}$ is given as:
\begin{equation}
    X_{\text{VG}} = f_{\Theta}(X).
    \label{eq:vqgan_relation}
\end{equation}

If $f_{\Theta}^{-1}$ represents the VQGAN decoder, the image can be reconstructed from the compressed latent space using 
\begin{equation}
    f_{\Theta}^{-1}(X_{\text{VG}}) \approx X.
    \label{eq:vqgan_decoder}
\end{equation}

Stage B is a LDM that learns to generate the latent space $X_{\text{VG}}$ given a highly compressed latent representation $Z$ of $X$. This compact embedding is obtained via the EfficientNet-V2 encoder \cite{tan2019efficientnet}. During the forward process in training, the latents $X_{\text{VG}}$ are noised according to the following relation:
\begin{equation}
    X_{\text{VG},t} = \sqrt{\bar{\alpha}_t} \cdot X_{\text{VG},t} + \sqrt{1 - \bar{\alpha}_t} \cdot \epsilon.
    \label{eq:diffusion_relation}
\end{equation}
Here, $\bar{\alpha}_t$ specifies the noise schedule whereas $\epsilon$ is the noise sampled from a standard normal distribution $N(0,1)$. At any time-step $t$, with noised latents $X_{\text{VG},t}$ and noisy conditional embedding $\hat{Z}$, the LDM is trained to predict the noise $\bar{\epsilon}(X_{\text{VG},t}, t, \hat{Z})$. The training objective is to minimize the loss function $L$, defined as the Mean-Squared Error (MSE) between the predicted and actual noise:
\begin{equation}
    L = \mathbb{E}_{(X_{\text{VG},t}, t, \hat{Z}, \epsilon)} \left[\| \epsilon - \bar{\epsilon}(X_{\text{VG},t}, t, \hat{Z}) \|_2^2 \right].
    \label{eq:mse_loss}
\end{equation}
\indent Text embedding is also used as conditioning for Stage B in the original SC paper \cite{pernias2023wurstchen}. However, as noted in the paper itself, it has no significant impact on the reconstruction quality of stage B as the conditioning provided by the image embedding is much stronger. Thus, we do not consider text conditioning in our model. The fine-tuning of Stage B conditioned on $\hat{Z}$ makes it robust to channel impairments. Moreover, we do not consider stage C either as it is primarily responsible for text-to-image generation.

\section{System Model} \cref{fig:System_Model} shows the three phases of our system model i.e. semantic information extraction at the transmitter, noisy channel transmission, and image reconstruction at the receiver.

\subsection{Semantic Feature Extraction}
As in \cite{pernias2023wurstchen}, we utilize the pretrained EfficientNet-V2 image encoder to extract a compact image embedding. An input RGB image $X \in \mathbb{R}^{N \times H \times W}$ is encoded into a compressed embedding $Z=\mathcal{E}(X)$\footnote{The dimensionalities of $X$; $N$ is the number of channels, i.e. 3 for RGB, and $H$ and $W$ stand for the height and width pixel resolution respectively.}. Despite its compact size, this embedding contains well-generalized feature representations that provide stronger guidance to the diffusion model as compared to text embeddings. As a result, the reconstructed image is very similar to the original one, with differences in fine details only. Although image generation based solely on text conditioning is highly efficient in terms of bandwidth, it may result in reconstructions that are semantically quite different from the source image \cite{nam2024language}. Furthermore, as compared to segmentation map-based conditioning \cite{grassucci2023generative}, image embeddings offer better reconstruction fidelity. Although segmentation maps retain spatial structure, they often lose crucial details such as texture, color, and fine-grained features. Additionally, because they provide only class-level information, the same segmentation map can yield multiple plausible reconstructions, introducing variability. To achieve reliable, predictable, and efficient SIC, we propose using rich image embedding as a more effective conditioning signal, ensuring reduced generation randomness and high-fidelity reconstruction of transmitted images.

\subsection{Communication Channel}
To maintain conformity with most previous works \cite{grassucci2023generative, yilmaz2024high, chen2024casc, yang2024rate}, we consider the widely adopted additive white Gaussian noise (AWGN) channel in our simulations. The extracted image embedding Z is transmitted across the AWGN channel where the noise $\epsilon$ is sampled from a zero-mean normal distribution $N(0,\sigma^2)$ with variance $\sigma^2$. If $P$ denotes the received signal power, the channel conditions are characterized by the Signal-to-Noise Ratio (SNR):
\begin{equation}
    \text{SNR} = 10 \log \left(\frac{P}{\sigma^2}\right) \, \text{(dB)}.
    \label{eq:SNR}
\end{equation}
Depending upon the SNR level, noise is added to $Z$ and the distorted embedding $\hat{Z}$ is obtained as
\begin{equation}
    \hat{Z} = Z + \epsilon.
    \label{eq:noisy_embedding}
\end{equation}

\subsection{Image Reconstruction}
The noisy image embedding $\hat{Z}$ is used as a conditioning signal to the diffusion model at the receiver side. It should be noted that in \cite{cicchetti2024language}, a text-conditioned diffusion model starts sampling from a noisy version of the image embedding, whereas, in our model, a significantly more compressed image embedding is used purely as a conditioning signal. After the conditional denoising process is complete, the output of the LDM is the predicted latent space $\hat{X}_{\text{VG}}$ where the VQGAN decoder operates. Finally, in accordance with \cref{eq:vqgan_decoder}, the generated image $\hat{X}$ is obtained using $f_{\Theta}^{-1}(\hat{X}_{\text{VG}}) = \hat{X}$.

\begin{figure*}[t]
    \centering
    \includegraphics[width=\textwidth]{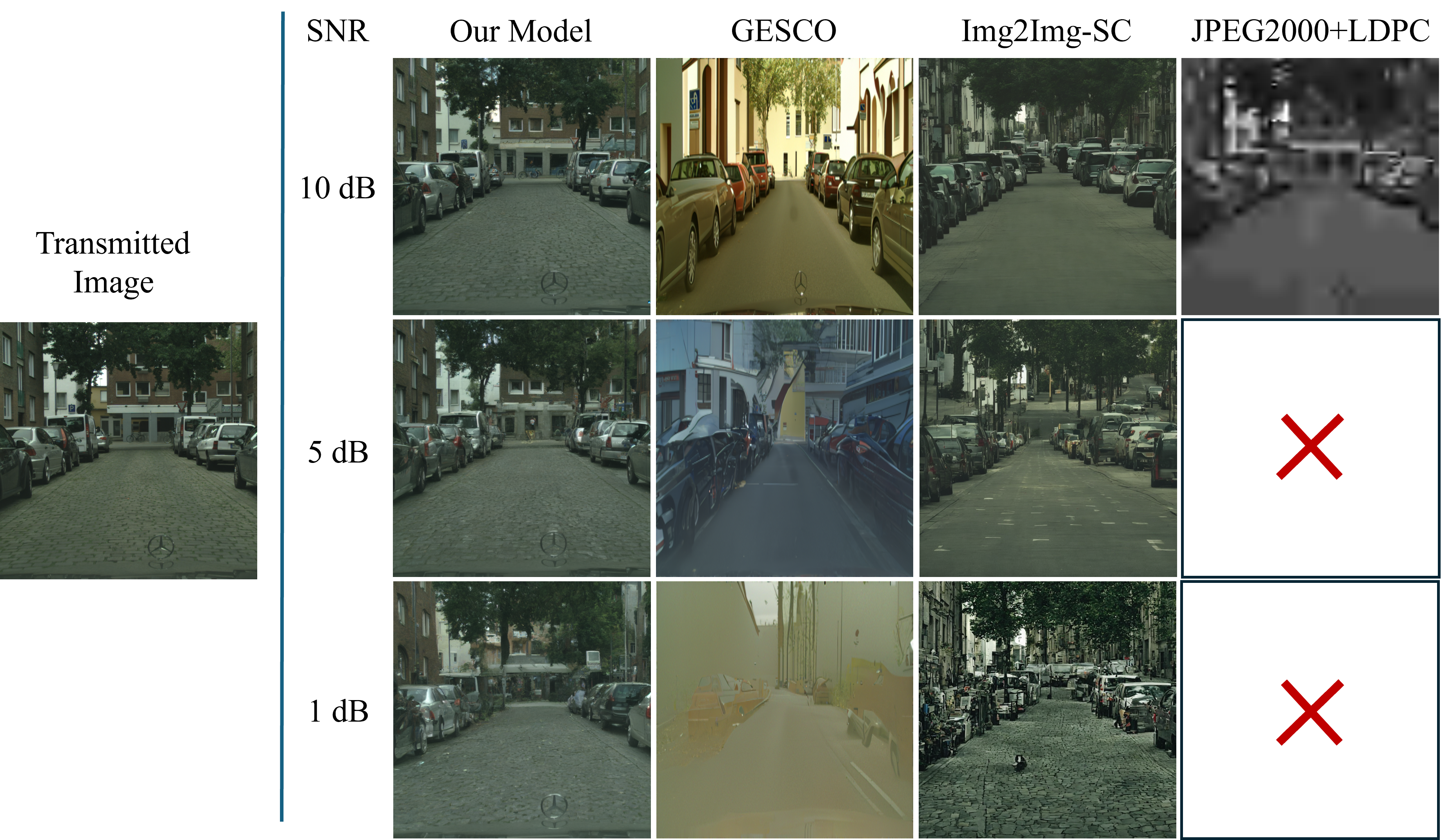}
    \caption{Image reconstructions using our model, GESCO, Img2Img-SC and JPEG2000$+$LDPC in low SNR conditions. It can be observed that our model generates the most semantically similar images with the least generation randomness. The red crosses indicate that the JPEG2000$+$LDPC system was unable to recover the image at the corresponding SNR.}
    \label{fig:generation_comparison}
\end{figure*}

\section{Experimental Evaluation}

\subsection{Model Training}
We train our model using the Cityscapes dataset \cite{cordts2016cityscapes}. The dataset contains $3000$ training, $500$ validation, and $1500$ test images. All images are resized to $1024\times1024$ resolution. We finetune the pre-trained stage B checkpoint for $15000$ steps using a batch size of 4, learning rate of $1\times10^{-4}$, and AdamW optimizer. To improve generalization and robustness, the SNR is randomly selected to be between $1-20$ dB. At each training step, image embeddings are extracted and transmitted across the AWGN channel. The model is trained to use the noisy embeddings as conditioning to reconstruct images with the objective of minimizing the MSE loss in accordance with \cref{eq:mse_loss}. In addition to the Cityscapes dataset, we also evaluate our model's performance on the DIV2K dataset \cite{Agustsson_2017_CVPR_Workshops}, which is composed of highly diverse images. We do not finetune our model again for this dataset to investigate how well it generalizes on completely different and unseen data. All the training and simulations have been performed using a single NVIDIA RTX A6000 (48-GB) GPU. All code scripts and fine-tuned model weights will be accessible at: \url{https://github.com/abilalk02/SC-SIC}.

\subsection{Simulation Settings}
We compare the performance of our model with (i) the diffusion SIC model conditioned on segmentation maps (GESCO) \cite{grassucci2023generative}, (ii) the Stable Diffusion-based SIC model that transmits text and image embeddings (Img2Img-SC) \cite{cicchetti2024language}, and (iii) the conventional JPEG2000 compression with Low-Density Parity-Check (LDPC) error correction approach. For evaluation, we generate $100$ samples using each model with channel SNR values of $1,5,10,15$ and $20$ dB respectively. All samples are of resolution $512\times512$, except for GESCO, where the resolution is $256\times512$\footnote{It was not possible to generate $512\times512$ images using GESCO without altering the model architecture.}. For sampling with GESCO and Img2Img-SC, $1000$ and $30$ denoising steps are used, respectively, as in the original papers. For JPEG2000$+$LDPC, Quadrature Amplitude Modulation (QAM) is used and the LDPC coding rate is set to $1/2$ following the method described in \cite{bourtsoulatze2019deep}. 

\textbf{Performance Metrics:} To evaluate the perceptual and semantic similarity between the original and generated images, we calculate the Learned Perceptual Image Patch Similarity (LPIPS) score \cite{zhang2018perceptual}, Fréchet Inception Distance (FID) score \cite{Seitzer2020FID} and Structural Similarity Index Measure (SSIM) \cite{wang2004image}. We also measure the Peak Signal-to-Noise Ratio (PSNR) to evaluate pixel-level similarity between images. Lower values of LPIPS and FID indicate better performance, whereas higher values of SSIM and PSNR indicate better performance.

\subsection{Results}

\subsubsection{Image Reconstruction Quality}
We first evaluate the reconstruction quality of our model against existing approaches, including GESCO, Img2Img-SC, and the JPEG2000$+$LDPC framework. \cref{fig:generation_comparison} shows the reconstruction of a transmitted image at the receiver end using the four models under low SNR conditions. Our model consistently achieves the most accurate reconstructions of the original image. Even at extremely low SNR levels of 5 dB and 1 dB, it preserves object clarity and recognizability. In contrast, the reconstruction quality of GESCO deteriorates rapidly as SNR decreases, leading to significant visual degradation. Moreover, the output produced by Img2Img-SC is loosely tied to the original image because text conditioning introduces significant randomness in the generation process. Finally, the conventional JPEG2000$+$LDPC produces heavily distorted output, and error correction completely fails at low SNR, as was observed earlier \cite{bourtsoulatze2019deep, Jiang2024}. For cases where it fails to reconstruct the images, we set the PSNR and SSIM scores to $0$, whereas LPIPS and FID scores are assigned an arbitrary maximum value of $1$ and $500$ respectively.

\begin{figure}[t]  
    \includegraphics[width=\columnwidth]{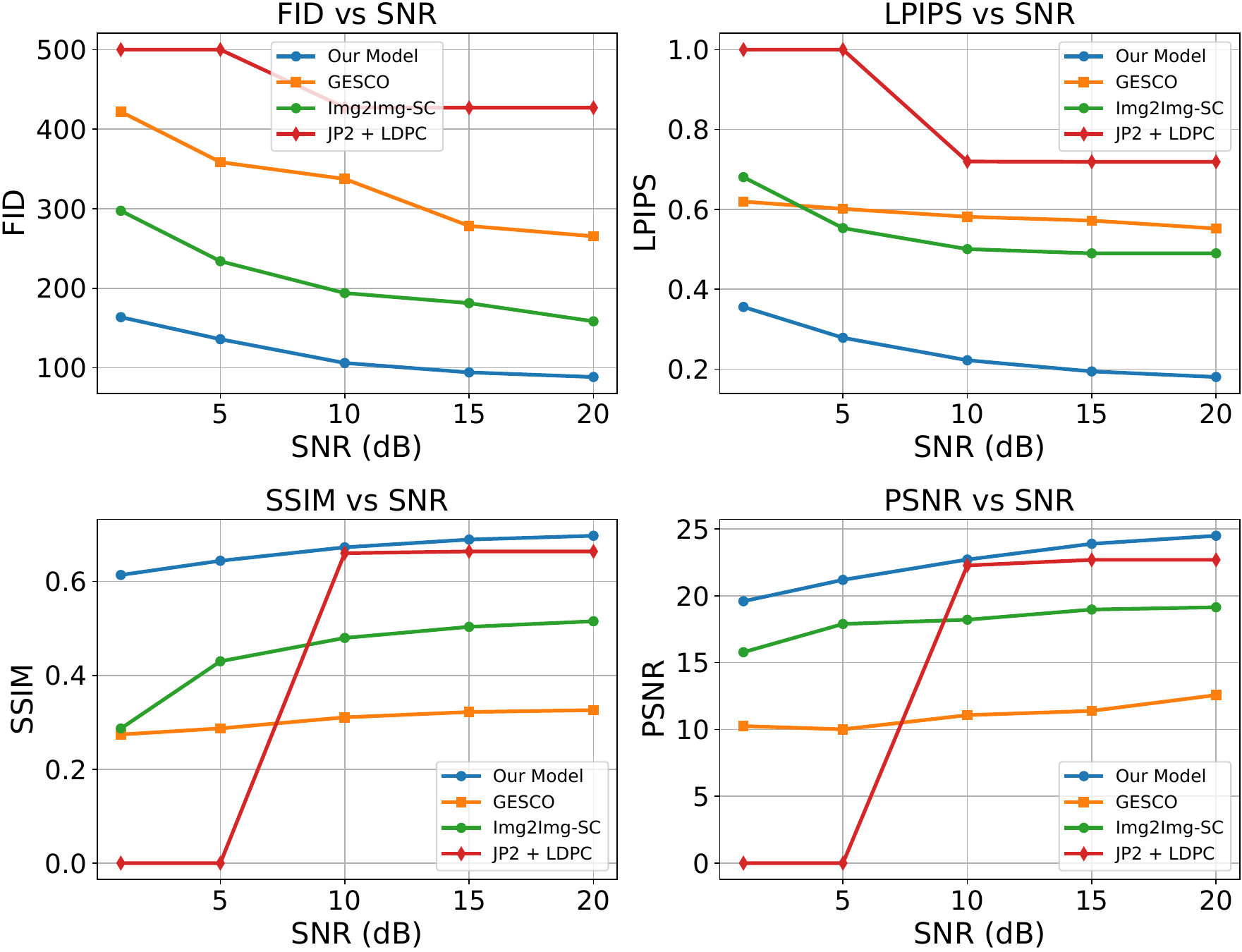}
    \vspace{-15pt}
    \caption{Performance comparison between our model, GESCO, Img2Img-SC and JP2$+$LDPC at different SNRs.}
    \label{fig:performance_metrics}
    %\vspace{-15pt}
    \vskip -0.15in
\end{figure}

\indent The comparison across performance metrics on the Cityscapes test data, shown in \cref{fig:performance_metrics}, also reveals that our model achieves the best results. In terms of FID and LPIPS, on average, our model improves on the results of the next-best approach from Img2Img-SC by $43\%$ and $55\%$, respectively. Similarly, in terms of SSIM and PSNR, our model gives the best results, maintaining good performance even at low SNR. For SNR greater than 10 dB, JPEG2000$+$LDPC achieves comparable PSNR and SSIM to our model even though its reconstructions are heavily distorted, have artifacts, and lack details. This can be attributed to the fact that JPEG2000 compression preserves low-frequency components and structural integrity. PSNR and SSIM primarily assess pixel-level accuracy and structural similarity, respectively. In contrast, LPIPS and FID are more sensitive to perceptually significant distortions, capturing the loss of fine details, reduced realism, and unnatural textures. Thus, high PSNR and SSIM scores can misleadingly overestimate the performance of JPEG2000$+$LDPC, failing to reflect the perceptual degradation. Moreover, as discussed, the conventional method fails to reconstruct the images at low SNR. Overall, our model improves SSIM by $56\%$ and PSNR by $23\%$ as compared to Img2Img-SC. The results of our model improve further when generating $1024\times1024$ images.

\subsubsection{Inference Speed and Bandwidth Efficiency}

\begin{figure}[t]
    \includegraphics[width=\columnwidth]{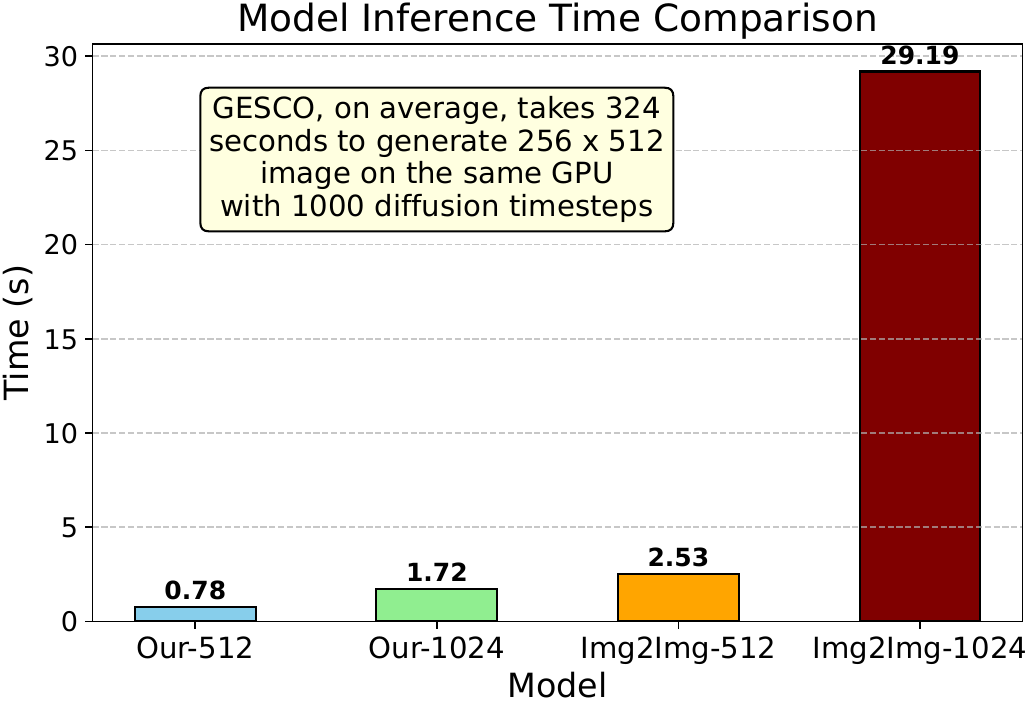}
    \vspace{-15pt}
    \caption{Inference time comparison of our model with GESCO and Img2Img-SC.}
    \label{fig:time_comparison}
    \vspace{-5pt}
\end{figure}

\indent In terms of computational complexity, we evaluate both inference latency and the dimensionality of the transmitted data. As shown in \cref{fig:time_comparison}, the model from \cite{grassucci2023generative}, which does not utilize an LDM, exhibits significantly higher latency, requiring $5$ minutes and $24$ seconds for image reconstruction with $T = 1000$ denoising steps. Our method achieves substantially lower inference time, just $0.78$ seconds for $512\times512$ images, making it $3\times$ faster than Img2Img-SC. For $1024\times1024$ images, our model accelerates reconstruction further, achieving speeds over $16\times$ faster than that of Img2Img-SC.

\vspace{-20pt}
\setlength{\extrarowheight}{5pt}
\begin{table}[h]
\caption{Dimensionality Comparison}
%\vskip 0.1in
\begin{center}
\resizebox{\linewidth}{!}{
\begin{tabular}{|c|c|c|c|}
\hline
\textbf{Transmitted Data} & \textbf{Dimensionality} & \textbf{Compression Ratio} & \textbf{\% of original} \\
\hline
Original Image & $[3,512,512]$ & $-$ & $-$ \\
\hline
Our Model & $[16,12,12]$ & $341$ & $0.29\%$ \\
\hline
Img2Img-SC & $[4,64,64]$ & $48$ & $2.08\%$ \\
\hline
DIFFSC & $[8,32,32]$ & $96$ & $1.04\%$ \\
\hline
CASC & $[8,32,32]$ & $96$ & $1.04\%$ \\
\hline
\end{tabular}
\label{table:dimensionality}
}
\end{center}
\vskip -0.1in
\end{table}

Moreover, in terms of dimensionality, \cref{table:dimensionality} shows that we achieve a higher Compression Ratio (CR) as compared to other state-of-the-art DM-based SIC systems. Following the definition in \cite{Jiang2024}, where CR is defined as the ratio of the input image's dimensionality to that of its encoded representation, our approach compresses an RGB image of size $[3, 512, 512]$ into a compact embedding of $[16, 12, 12]$, achieving an exceptional CR of $341$ -- meaning that the transmitted data is only $0.29$\% of the original image size. This highlights the remarkable bandwidth efficiency of our method.

\subsubsection{Reconstruction Predictability}

We assess reconstruction predictability across varying SNR conditions using the LPIPS metric. For each case, we simulate image transmission $25$ times with fixed parameters, computing the mean $(\mu)$ and standard deviation $(\sigma)$ of LPIPS scores across all pairwise comparisons of generated images. As shown in \cref{table:Predictability}, our model achieves the lowest average LPIPS score and standard deviation, $(\mu\pm\sigma)=(0.173\pm0.003)$ at SNR$=20$dB, indicating minimal generation randomness. Thus, the proposed model is able to reconstruct images reliably and consistently.

\begin{table}[h]
\caption{Predictability Comparison}
%\vskip 0.15in
\begin{center}
\resizebox{\linewidth}{!}{
\begin{tabular}{|c|c|c|c|c|}
\hline
\multirow{2}{*}{\textbf{SNR (dB)}} & \multicolumn{4}{c|}{\textbf{LPIPS Score ($\mu \pm \sigma$)}} \\
\cline{2-5}  % Horizontal line below LPIPS Score header
 & \textbf{Our-1024} & \textbf{Our-512} & \textbf{GESCO} & \textbf{Img2Img-SC} \\
\hline
$20$  & $0.173 \pm 0.003$ & $0.205 \pm 0.005$ & $0.401 \pm 0.014$ & $0.520 \pm 0.011$ \\
\hline
$15$  & $0.195 \pm 0.003$ & $0.223 \pm 0.006$ & $0.433 \pm 0.012$ & $0.541 \pm 0.017$ \\
\hline
$10$  & $0.229 \pm 0.003$ & $0.264 \pm 0.008$ & $0.424 \pm 0.017$ & $0.522 \pm 0.012$ \\
\hline
$5$   & $0.287 \pm 0.004$ & $0.314 \pm 0.009$ & $0.575 \pm 0.021$ & $0.554 \pm 0.019$ \\
\hline
$1$   & $0.351 \pm 0.006$ & $0.371 \pm 0.013$ & $0.613 \pm 0.017$ & $0.578 \pm 0.019$ \\
\hline
\end{tabular}
\label{table:Predictability}
}
\end{center}
\vskip -0.1in
\end{table}

\begin{figure}[t]  
    \includegraphics[width=\columnwidth]{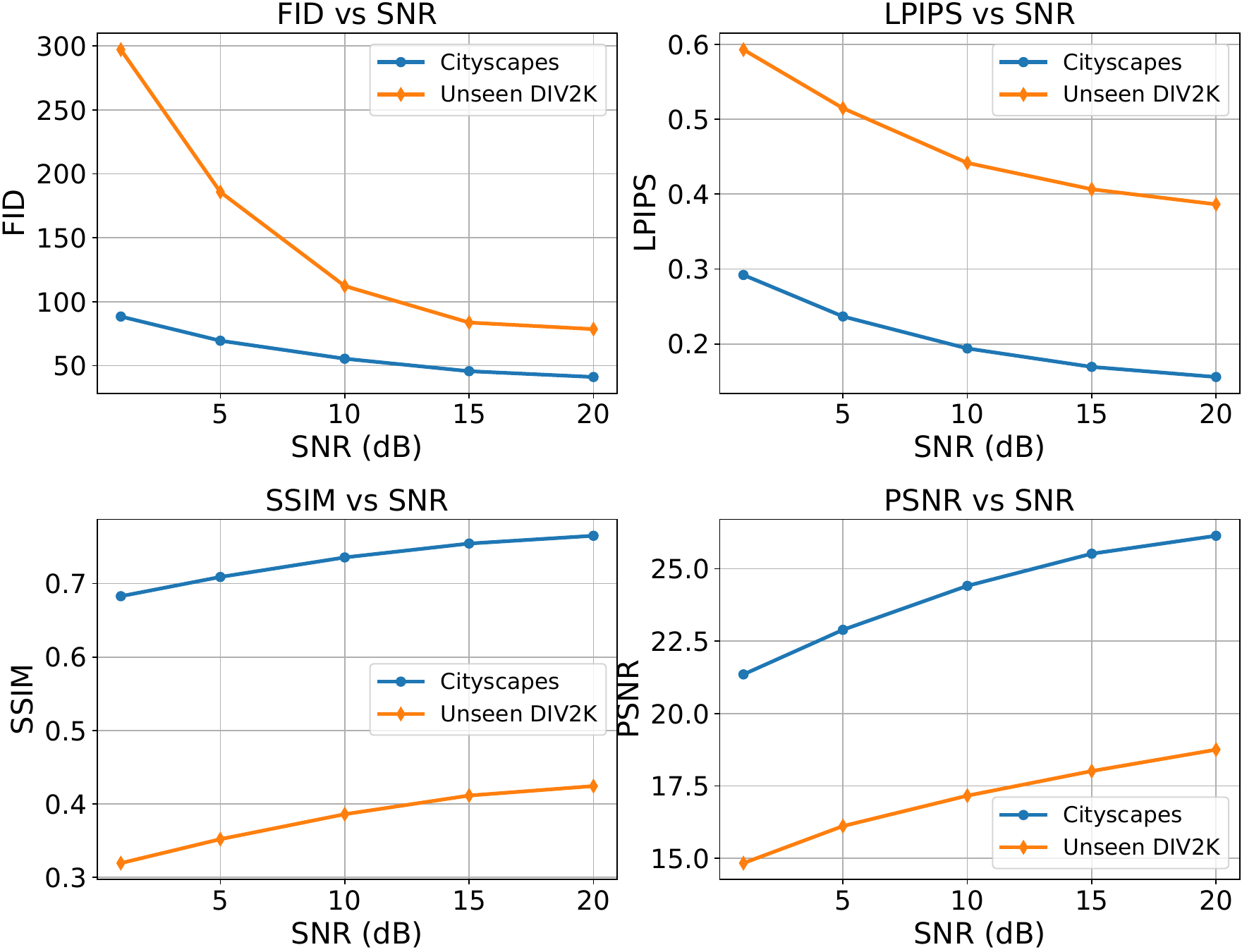}
    \vspace{-15pt}
    \caption{Performance of our model on unseen DIV2K data.}
    \label{fig:unseen_data}
    %\vspace{-15pt}
    %\vskip -0.15in
\end{figure}

\begin{figure}[t]  
    \includegraphics[width=\columnwidth]{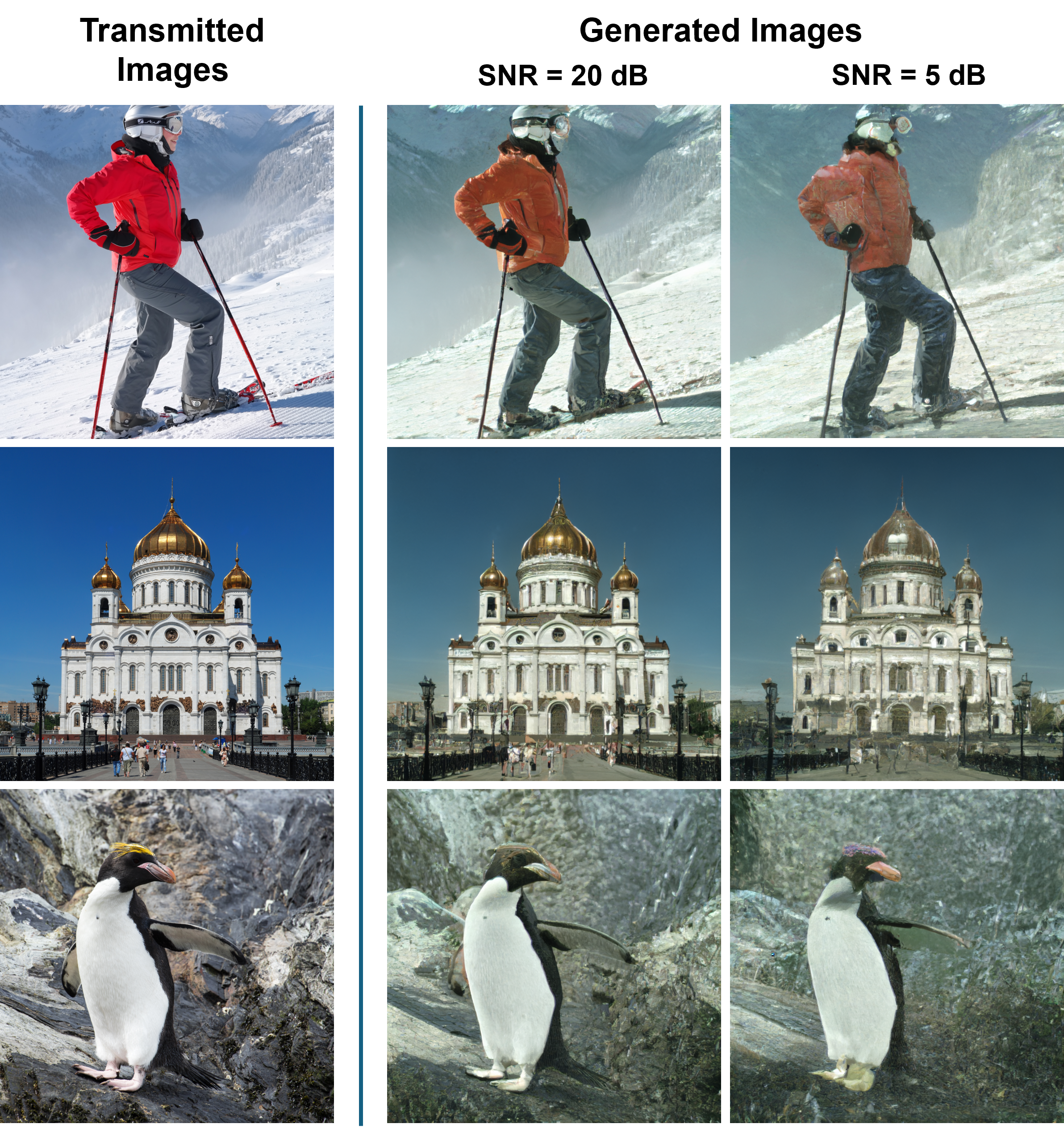}
    \vspace{-15pt}
    \caption{Image reconstructions on unseen DIV2K data. It can be seen that the model does well to mitigate the noise and reconstruct semantically similar images considering that it was not finetuned for this dataset.}
    \label{fig:div2k_reconstructions}
    %\vspace{-15pt}
    %\vskip -0.15in
\end{figure}

\subsubsection{Generalization on Unseen Data}

We also analyze the performance of our model, trained on the Cityscapes dataset, on entirely unseen data. For this purpose, we use the DIV2K dataset that contains diverse images, including landscapes, people, architecture, and animals. \cref{fig:unseen_data} indicates that there is a significant degradation in performance on this new data across all four metrics. For example, at an SNR of $15$ dB, LPIPS increases from 0.17 to 0.4, whereas FID increases from 45 to 83, indicating a substantial loss in perceptual quality. However, a closer look at the generated images, \cref{fig:div2k_reconstructions}, reveals that much of this degradation may be attributed to the sharp differences in the colors between the original and generated images. The model does fairly well to reconstruct these unseen images and mitigate the effects of noise, but since it is finetuned on the Cityscapes dataset, the generated images have a color tone that resembles very closely to that of the images in the said dataset. These results suggest that fine-tuning a Stable Cascade model on a single large and highly diverse dataset may enable it to handle a wide range of image types with strong performance.

\subsubsection{Ablation Studies}

\begin{figure*}[t]
    \centering
    \subfloat[]{%
        \centering
        \includegraphics[width=0.48\textwidth]{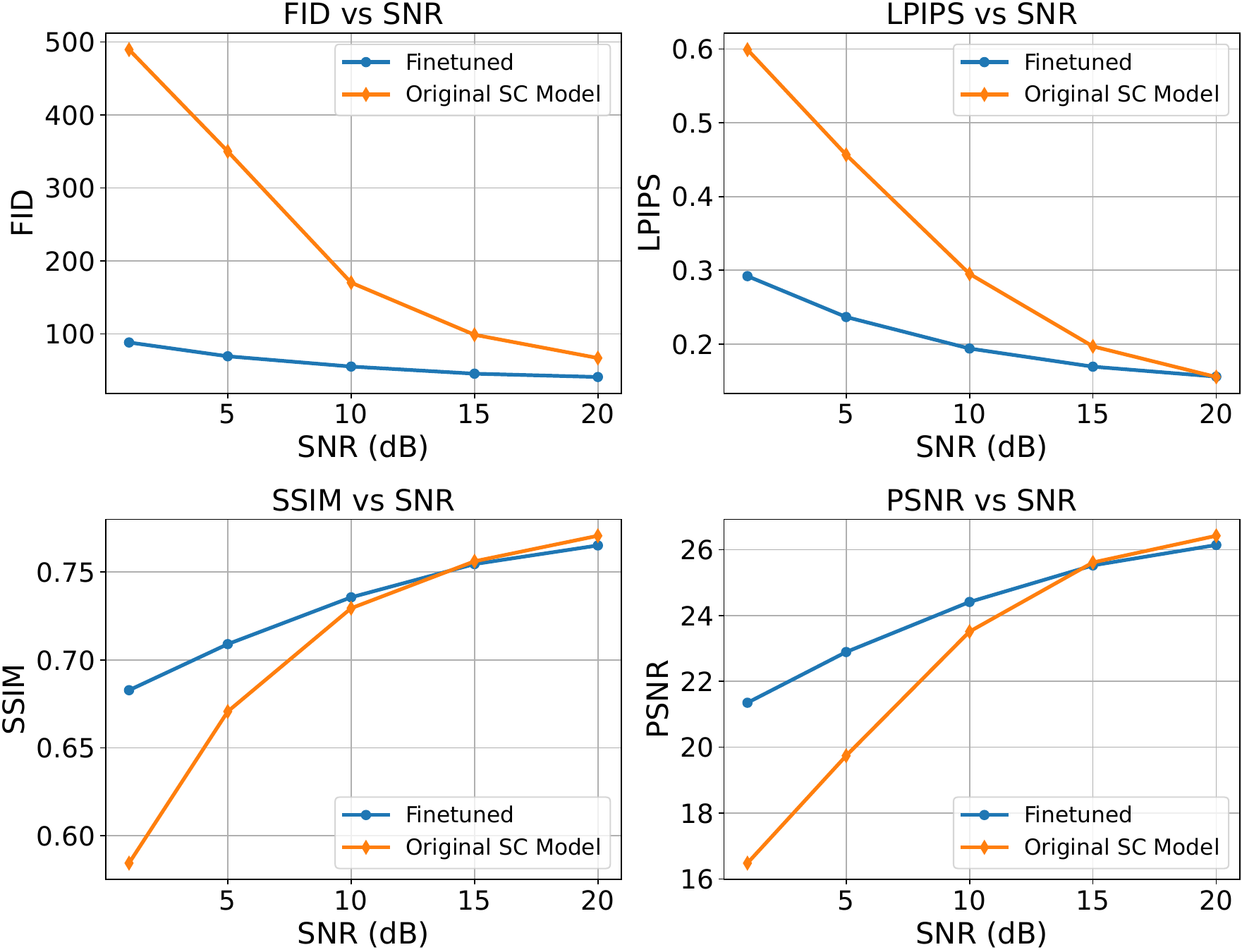}
        \label{fig:ft_vs_original}
    } \hfill
    \subfloat[]{%
        \centering
        \includegraphics[width=0.48\textwidth]{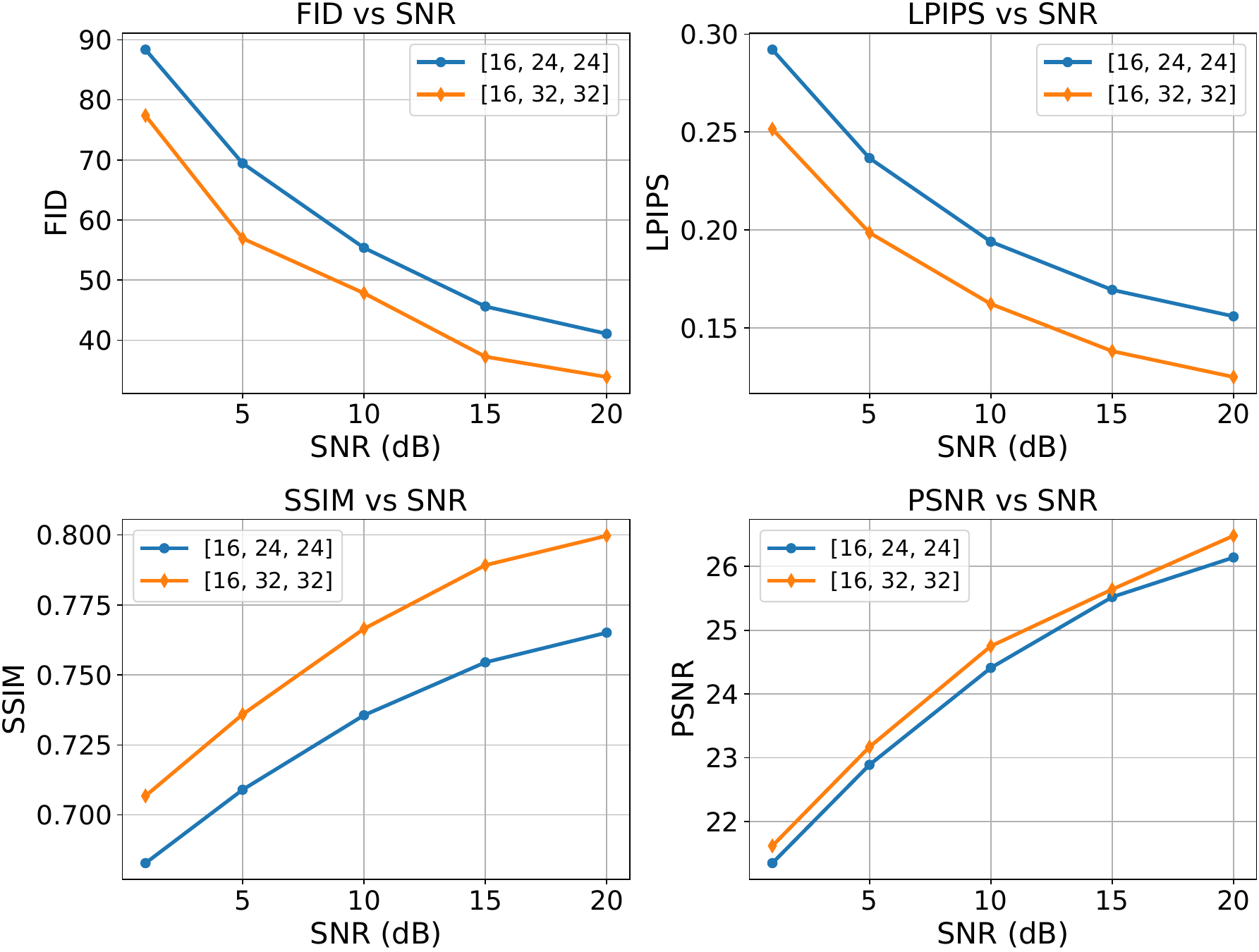}
        \label{fig:embedding_size}
    }
    \caption{Results of ablation experiments highlighting (a) the performance gains obtained via fine-tuning and (b) the impact of increasing the embedding size from [16, 24, 24] to [16, 32, 32] on performance metrics.}
    \label{fig:ablation}
\end{figure*}

\begin{figure}[t]  
    \includegraphics[width=\columnwidth]{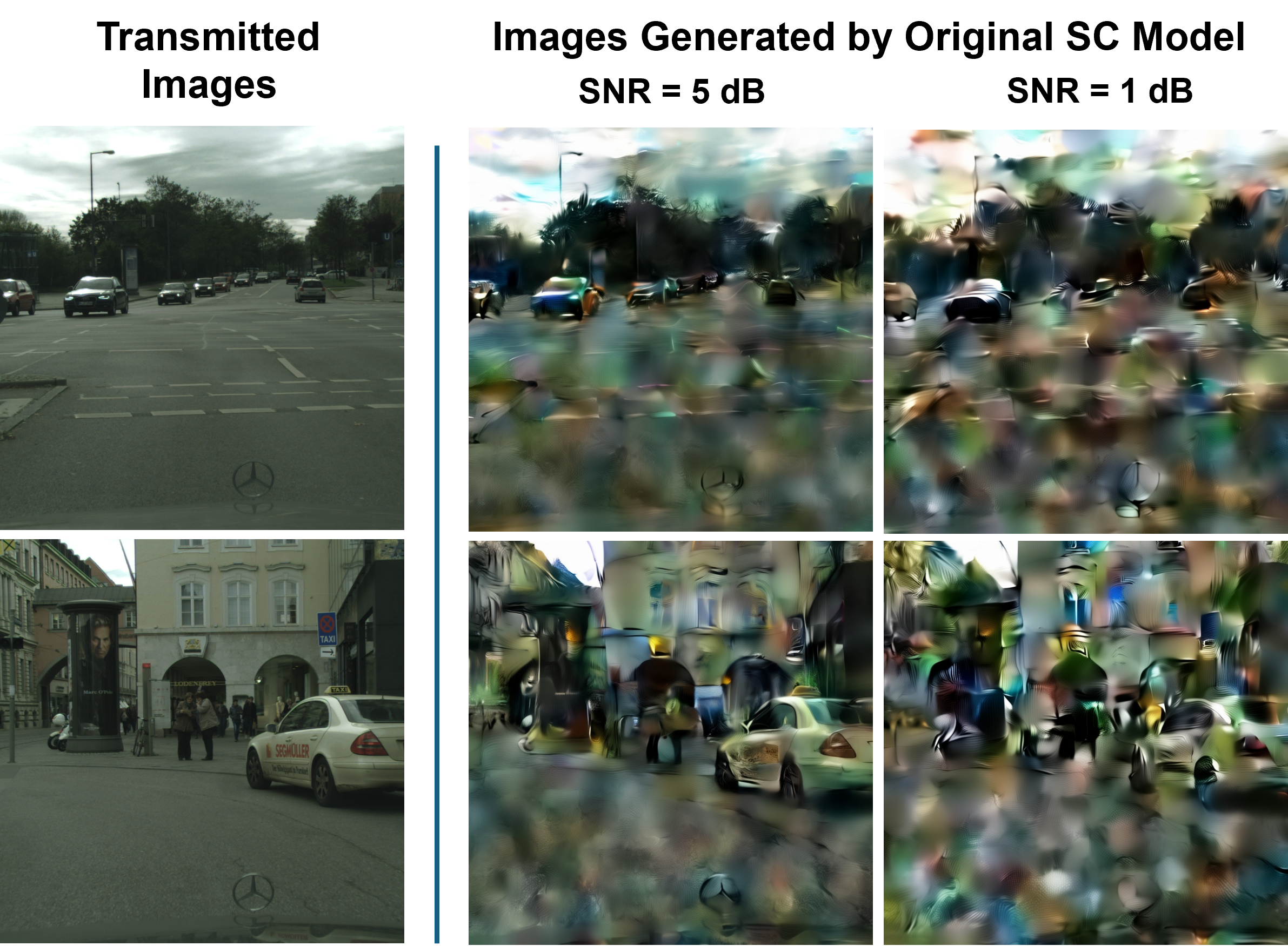}
    \vspace{-15pt}
    \caption{Images reconstructed by the original Stable Cascade model. It can be seen that without proper fine-tuning, the original model fails to deal with the effects of channel noise.}
    \label{fig:original_recons}
    %\vspace{-15pt}
    \vskip -0.15in
\end{figure}

Finally, we perform ablation tests to compare the performance of our fine-tuned model against the original Stable Cascade model in the semantic image communication scenario. \cref{fig:ft_vs_original} shows that without fine-tuning, the original model’s performance degrades sharply with decreasing SNR. In particular, at SNR less than $10$ dB, the images generated using the original model are heavily corrupted by noise. This is also evident from \cref{fig:original_recons}, which shows that the original model is unable to mitigate the channel effects. These findings validate our training approach and demonstrate the substantial performance gains achieved by fine-tuning the model to work with noisy image embedding as a conditioning signal. 

We also analyze the impact of increasing the size of the extracted image embedding on the generation quality for $1024\times1024$ images. It can be seen from \cref{fig:embedding_size} that there is a noticeable improvement in performance across all four performance metrics when the embedding size is increased from $[16, 24, 24]$ to $[16, 32, 32]$. Quantitatively, on average, LPIPS, FID, and SSIM scores improve by greater than $10\%$. However, these improvements come at a cost to the compression ratio that drops from $341$ to $192$. Hence, there is an understandable tradeoff between performance and bandwidth efficiency

\section{Conclusion}

In this paper, we introduce a novel DM-based SIC framework that leverages the Stable Cascade architecture to achieve an exceptional balance of speed, compression, and fidelity under noisy channel conditions. Our method transmits a highly compact image embedding, only $0.29$\% of the original size, and reconstructs $512\times512$ images in just $0.78$ seconds -- $3\times$ faster than Img2Img-SC. Extensive evaluations using perceptual quality metrics, including LPIPS, SSIM, and FID, demonstrate the noise robustness of our approach and its superiority over existing benchmarks. Additionally, our framework minimizes generation randomness by achieving an LPIPS score variance of only 0.003 at SNR greater than $10$dB, ensuring faithful and consistent image reconstruction. Future work may explore further optimizations to minimize inference time and extend the framework to high-fidelity semantic video communication.

\section*{Acknowledgements}

This research has received funding from the European Union's Horizon Europe research and innovation programme MSCA-DN NESTOR (G.A. 101119983). The authors also acknowledge EPSRC project TRANSNET (EP/R035342/1). Experiments were run on Aston EPS Machine Learning Server, funded by the EPSRC Core Equipment Fund, Grant EP/V036106/1.

%\bibliography{references}
\bibliographystyle{icml2025}

%%%%%%%%%%%%%%%%%%%%%%%%%%%%%%%%%%%%%%%%%%%%%%%%%%%%%%%%%%%%%%%%%%%%%%%%%%%%%%%
%%%%%%%%%%%%%%%%%%%%%%%%%%%%%%%%%%%%%%%%%%%%%%%%%%%%%%%%%%%%%%%%%%%%%%%%%%%%%%%
% APPENDIX
%%%%%%%%%%%%%%%%%%%%%%%%%%%%%%%%%%%%%%%%%%%%%%%%%%%%%%%%%%%%%%%%%%%%%%%%%%%%%%%
%%%%%%%%%%%%%%%%%%%%%%%%%%%%%%%%%%%%%%%%%%%%%%%%%%%%%%%%%%%%%%%%%%%%%%%%%%%%%%%
%\newpage
%\appendix
%\onecolumn
%\section{You \emph{can} have an appendix here.}

%%%%%%%%%%%%%%%%%%%%%%%%%%%%%%%%%%%%%%%%%%%%%%%%%%%%%%%%%%%%%%%%%%%%%%%%%%%%%%%
%%%%%%%%%%%%%%%%%%%%%%%%%%%%%%%%%%%%%%%%%%%%%%%%%%%%%%%%%%%%%%%%%%%%%%%%%%%%%%%

\end{document}